\documentclass[twocolumn]{webofc}

\usepackage[varg]{txfonts}   % Web of Conferences font
\usepackage{hyperref}
\usepackage{url}
\hypersetup{colorlinks=true,citecolor=blue,urlcolor=blue,linkcolor=blue}
\begin{document}
\title{Analysis of RF Sheath-Driven Tungsten Erosion at RF Antenna in the WEST Tokamak }

\author{
    \firstname{A.} \lastname{Kumar}\inst{1}\fnsep\thanks{\email{kumara@ornl.gov}} \and
    \firstname{W.} \lastname{Tierens}\inst{1} \and
    \firstname{T.} \lastname{Younkin}\inst{1} \and
    \firstname{C.} \lastname{Johnson}\inst{1} \and
    \firstname{C.} \lastname{Klepper}\inst{1} \and
    \firstname{A.} \lastname{Diaw}\inst{1} \and
    \firstname{J.} \lastname{Lore}\inst{1} \and
    \firstname{A.} \lastname{Grosjean}\inst{2} \and
    \firstname{G.} \lastname{Urbanczyk}\inst{3} \and
    \firstname{J.} \lastname{Hillairet}\inst{4} \and
    \firstname{P.} \lastname{Tamain}\inst{4} \and
    \firstname{L.} \lastname{Colas}\inst{4} \and
    \firstname{C.} \lastname{Guillemaut}\inst{4} \and
    \firstname{D.} \lastname{Curreli}\inst{5} \and
    \firstname{S.} \lastname{Shiraiwa}\inst{6} \and
    \firstname{N.} \lastname{Bertelli}\inst{6} \and
    \firstname{The} \lastname{WEST Team}\inst{7}
}

\institute{
    Oak Ridge National Laboratory, 1 Bethel Valley Road, Oak Ridge, TN-37831, USA
    \and
    University of Tennessee, Knoxville, TN 37996, USA
    \and
    Institut Jean Lamour UMR 7198 CNRS - Université de Lorraine, 2 allée André Guinier, F-54011 Nancy, France
    \and
    CEA, IRFM, F-13108 Saint-Paul-lez-Durance, France
    \and
    University of Illinois at Urbana-Champaign, Urbana, IL 61801, USA
    \and
    Princeton Plasma Physics Laboratory, Princeton, NJ 08536, USA
    \and
    \url{http://west.cea.fr/WESTteam}
}
\abstract{
This study applies the newly developed STRIPE (Simulated Transport of RF Impurity Production and Emission) framework to interpret tungsten (W) erosion at RF antenna structures in the WEST tokamak. STRIPE integrates SolEdge3x for edge plasma backgrounds, COMSOL for 3D RF sheath potentials, RustBCA for sputtering yields, and GITR for impurity transport and ion energy–angle distributions. In contrast to prior work by Kumar et al. (2025) Nuclear Fusion, 65, 076039, which focused on framework validation for WEST ICRH discharge \#57877, the present study provides a spatially resolved analysis of gross W erosion at both Q2 antenna limiters under ohmic and ICRH conditions. Using 2D SolEdge3x profiles in COMSOL, STRIPE captures rectified sheath potentials exceeding 300~V, leading to strong upper-limiter localization. Both poloidal and toroidal asymmetries are observed and attributed to RF sheath effects, with modeled erosion patterns deviating from experiment—highlighting sensitivity to sheath geometry and plasma resolution. High-charge-state oxygen ions (O$^{6+}$–O$^{8+}$) dominate erosion, while D$^+$ contributes negligibly. A plasma composition of 1\% oxygen and 98\% deuterium is assumed. STRIPE predicts a 30-fold increase in gross W erosion from ohmic to ICRH phases, consistent with W-I 400.9~nm brightness measurements. Agreement within 5\% (ohmic) and 30\% (ICRH) demonstrates predictive capability and supports STRIPE’s application in reactor-scale antenna design.
}
\maketitle
\section{Introduction}
\label{sec:1}

Radio-frequency (RF) waves are a primary tool for heating and current drive in magnetically confined fusion (MCF) plasmas \cite{Maslov_2023}. With the move toward reactor-scale devices, such as SPARC \cite{Creely_2020}, RF-based systems are expected to serve as primary heating sources, further motivating the need to understand associated edge plasma effects.\footnotetext{This manuscript has been authored by UT-Battelle, LLC, under contract DE-AC05-00OR22725 with the US Department of Energy (DOE). The US government retains and the publisher, by accepting the article for publication, acknowledges that the US government retains a nonexclusive, paid-up, irrevocable, worldwide license to publish or reproduce the published form of this manuscript, or allow others to do so, for US government purposes. DOE will provide public access to these results of federally sponsored research in accordance with the DOE Public Access Plan (http://energy.gov/downloads/doe-public-access-plan).}  One major concern in high-power RF operation is the formation of rectified RF sheaths at plasma-facing components (PFCs), particularly near ICRH antenna structures. Driven by RF waves, these sheaths can sustain potentials of several hundred volts in existing devices \cite{Tierens_2024} and are projected to exceed 1 kV in future systems such as ITER \cite{colas:2025}. Sheath-accelerated ions, including fuel species and light impurities, can induce physical sputtering of antenna materials—typically high-Z metals like tungsten (W). The resulting impurity sources can enter the scrape-off layer (SOL), and under certain conditions, migrate toward the core plasma. This is especially problematic in H-mode regimes, where longer confinement times enhance impurity retention and degrade overall performance. 

Localized impurity generation driven by RF sheath rectification has been observed in toroidal devices~\cite{Colas:2022}. While mitigation strategies such as changing antenna power balance ratio and the strap phasing~\cite{Bobkov:2017}, wall conditioning~\cite{Waelbroeck_1989}, and field-aligned launchers~\cite{wukitch:2013} have shown limited success. Moreover, only a few initial efforts have been made to  modeled RF-induced PMI in detail~\cite{Klepper:2022, Zhang:2022}. Nevertheless, the fusion community continues to seek predictive, integrated tools capable of linking RF field structure, edge plasma conditions, and impurity sourcing.

In this work, the newly developed STRIPE (Simulated Transport of RF Impurity Production and Emission) framework \cite{kumar:2025} is applied to assess W-erosion from the ICRH antenna limiters in the WEST tokamak for  a dedicated L-mode ICRH discharge (\#57877). WEST, with its all-W-PFCs and recently demonstrated record long-pulse operation capability exceeding 1300 seconds and injected energy of 2~GJ~\cite{Dumont:2025}, provides a relevant platform for validating high-fidelity plasma-material interaction (PMI) models under reactor-relevant conditions. 

In contrast to previous work on this discharge~\cite{kumar:2025}, which focused on validating STRIPE against localized W emission near the antenna region, this study presents a spatially resolved, in-depth analysis of plasma conditions and gross W-erosion to address experimentally observed erosion hot spots with both toroidal and poloidal asymmetries at the right and left antenna limiters.   While STRIPE also supports modeling of redeposition, self-sputtering, and net erosion, those aspects are deferred to a companion study focused on whole-device transport of W-impurities originating from the ICRH antenna.

The remainder of this paper is structured as follows: Section~\ref{sec:2} introduces the STRIPE framework and  discusses main inputs to STRIPE framework. Section~\ref{sec:4}  presents results comparing thermal and RF sheath-driven erosion at the WEST ICRH limiters.  Conclusions are summarized in Section~\ref{sec:6}.

\section{Overview and Inputs of the STRIPE Framework}
\label{sec:2}

{STRIPE} (Simulated Transport of RF Impurity Production and Emission)~\cite{kumar:2025} is an integrated, multi-physics framework developed to model impurity-driven W erosion at ICRH antenna structures. It couples 2D edge plasma backgrounds from {SolEdge3x}~\cite{Bufferand_2015}, rectified RF sheath potentials from COMSOL~\cite{Tierens_2024}, ion trajectory and impact dynamics from {GITR}~\cite{Younkin:2021} and {RustBCA}~\cite{RustBCA}, and post-processing via {ColRadPy}~\cite{curt2019} to convert sputtered fluxes into spectroscopically observable emission.

\begin{figure}[h]
    \centering
    \includegraphics[width=\linewidth]{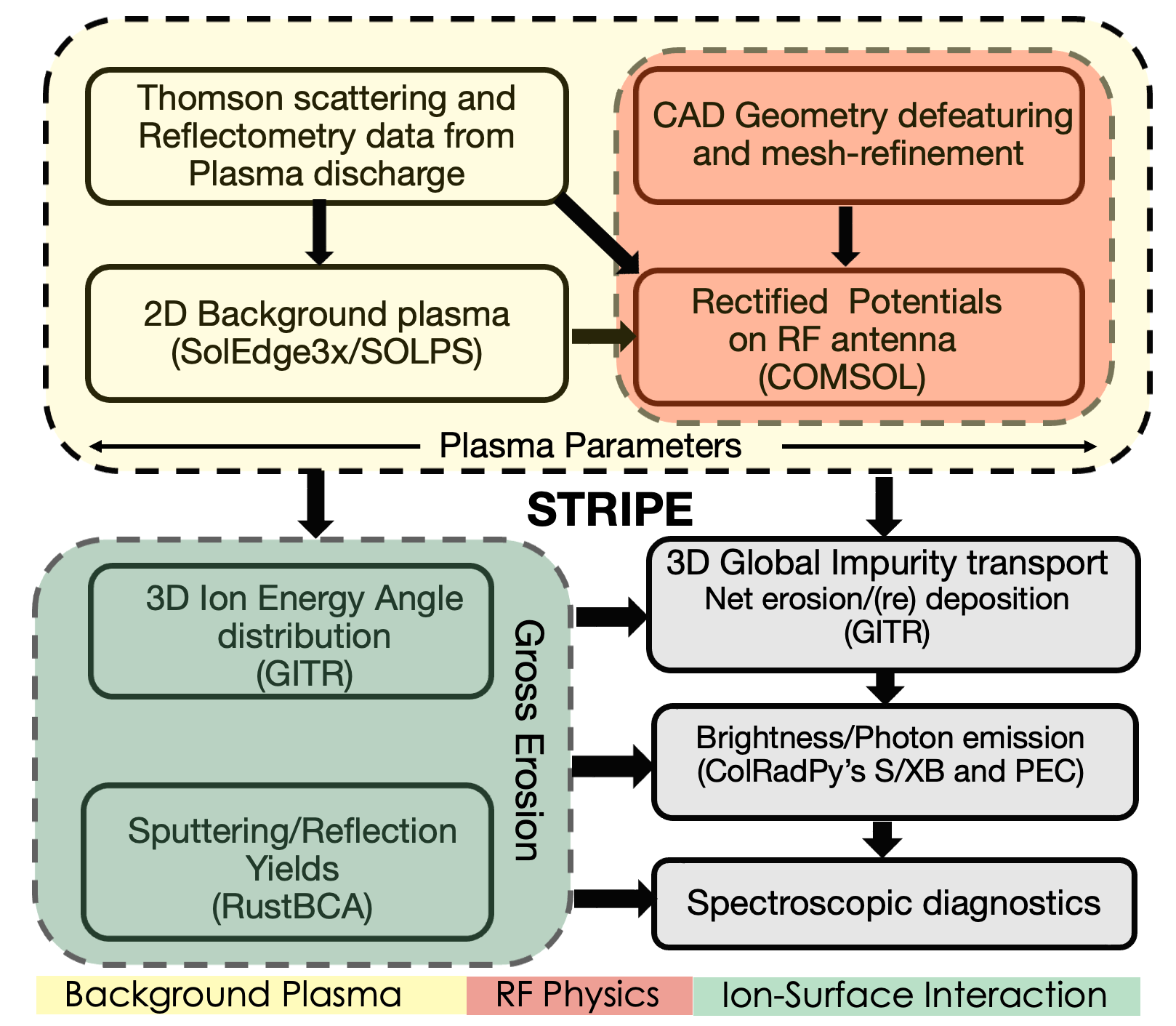}
    \caption{STRIPE workflow for modeling W erosion and impurity transport originating from ICRH antenna structures, coupling SolEdge3x, COMSOL, RustBCA, GITR, and ColRadPy.}
    \label{fig:1}
\end{figure}

The STRIPE workflow (Fig.~\ref{fig:1}) enables predictive modeling of RF-enhanced PMI by integrating sheath fields, charge-resolved impurity fluxes, and energy–angle-dependent sputtering yields. In this study, STRIPE is applied to model W erosion at the WEST Q2 ICRH antenna for discharge \#57877 under both ICRH and ohmic conditions.

The framework relies on spatially resolved physics inputs that capture gradients in plasma parameters, impurity species, and sheath voltages across 3D limiter geometries. These include SolEdge3x-derived electron density and temperature profiles, COMSOL-calculated sheath fields, and surface interaction data from RustBCA. The following subsections present these key inputs and how they are incorporated to simulate limiter-targeted impurity production and erosion.

\subsection{Limiter-Resolved Plasma Profiles and Charge State Distribution}
\label{subsec:31}

\begin{figure}[h]
    \centering
    \includegraphics[width=\linewidth]{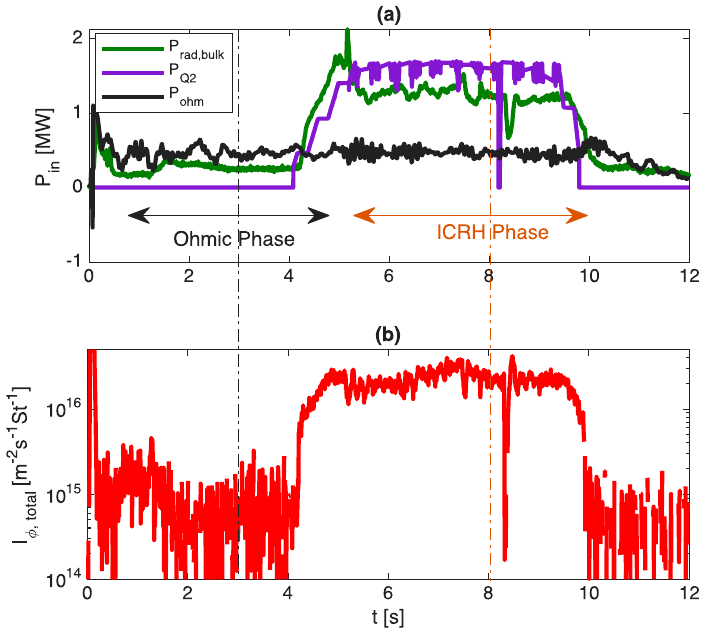}
    \caption{
    Time evolution of (a) heating power from ohmic and Q2 ICRH system, and radiated power; (c) total W-I 400.9~nm brightness near the ICRH antenna limiters. The transition from ohmic to ICRH leads to a factor of 25 increase in emission.
    }
    \label{fig:2}
\end{figure}

Figure~\ref{fig:2} shows the time evolution of key plasma parameters for WEST discharge \#57877, which serves as the experimental basis for the SolEdge3x simulations used in this study. The discharge begins with ohmic heating from \(t = 0\) to \(5\)~s and transitions to ICRH at \(t = 5\)~s, with a total power input of approximately 2.3~MW, including 1.7~MW delivered through the Q2 antenna. Panel (b) shows the visible brightness of the W I brightness averaged over all the antenna limiter lines of sight  which increases by more than a factor of 25 during the ICRH phase—highlighting the strong impact of RF power on impurity generation and motivating this erosion analysis.

Background SOL plasma conditions are reconstructed using wide-grid SolEdge3x–EIRENE simulations at two representative time points: \(t = 3~\text{s}\) (ohmic phase) and \(t = 8~\text{s}\) (ICRH phase). The simulation setup and boundary conditions are described in detail in~\cite{kumar:2025}. These 2D edge plasma solutions span the entire SOL and provide spatially resolved plasma parameters adjacent to the antenna-facing surfaces.

In {SolEdge3x}, the wall geometry is assumed toroidally symmetric, whereas the actual WEST ICRH antenna exhibits a distinct poloidal asymmetry about the midplane. This mismatch may reduce the fidelity of local plasma representation near the antenna and contribute to discrepancies between STRIPE predictions and experimental impurity emission. Drifts are disabled, and identical transport coefficients are applied for both ohmic and ICRH phases, enabling isolation of RF sheath rectification effects on tungsten erosion. Moreover, RF sheath rectification can itself modify local transport—for example, through $\mathbf{E} \times \mathbf{B}$ convective cells driven by gradients in the rectified DC potential. Nevertheless, by holding background transport fixed, impurity sourcing can be more directly attributed to variations in sheath potential rather than to changes in radial transport.

\begin{figure}[htbp]
\centering
\includegraphics[width=\linewidth]{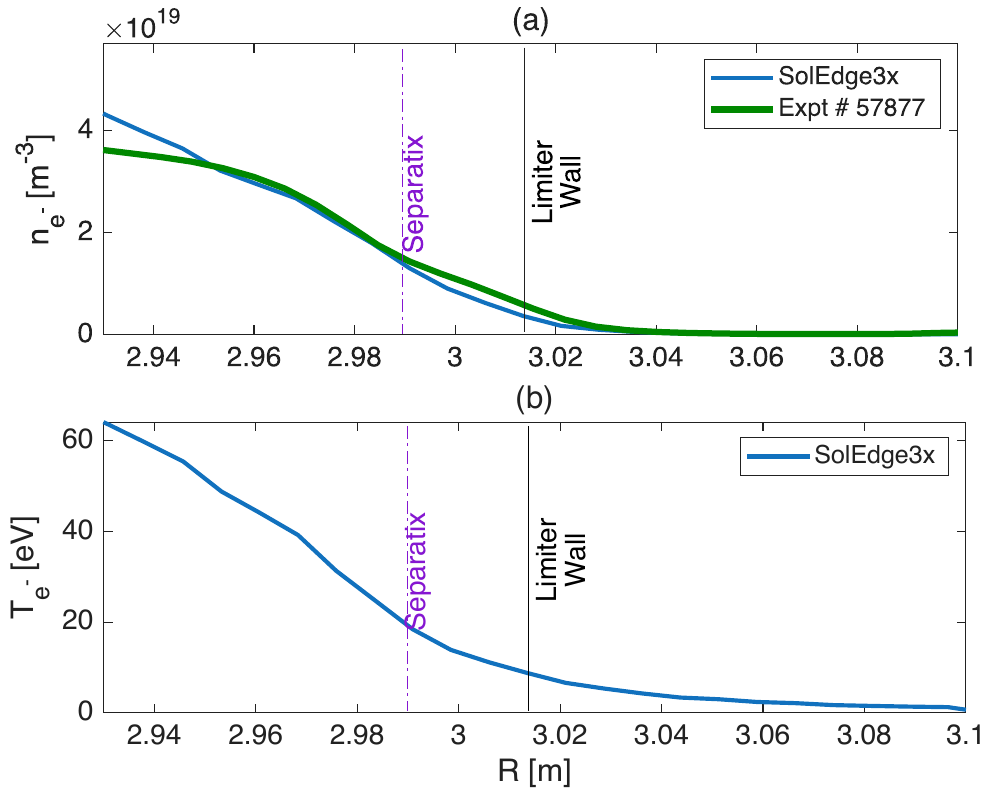}
\caption{Radial profiles of (a) electron density $n_{e^-}$ and (b) electron temperature $T_{e^-}$ at the OMP SOL from SolEdge3x simulations. The $n_{e^-}$ profile is compared with reflectometry data from WEST discharge \#57877, showing reasonably good agreement in SOL and far-SOL region.}
\label{fig:west_profiles}
\end{figure}

Figure~\ref{fig:west_profiles} compares the SolEdge3x-predicted electron density (\(n_{e^-}\)) with outer midplane (OMP) reflectometry measurements for WEST discharge \#57877. The simulation shows good agreement with experimental data in the radial region of interest. The corresponding simulated electron temperature (\(T_{e^-}\)) profile is also shown. However, no reliable \(T_{e^-}\) measurements are available during the ICRH phase,  and no  data were collected during the ohmic phase.

\begin{figure}[h]
    \centering
    \includegraphics[width=\linewidth]{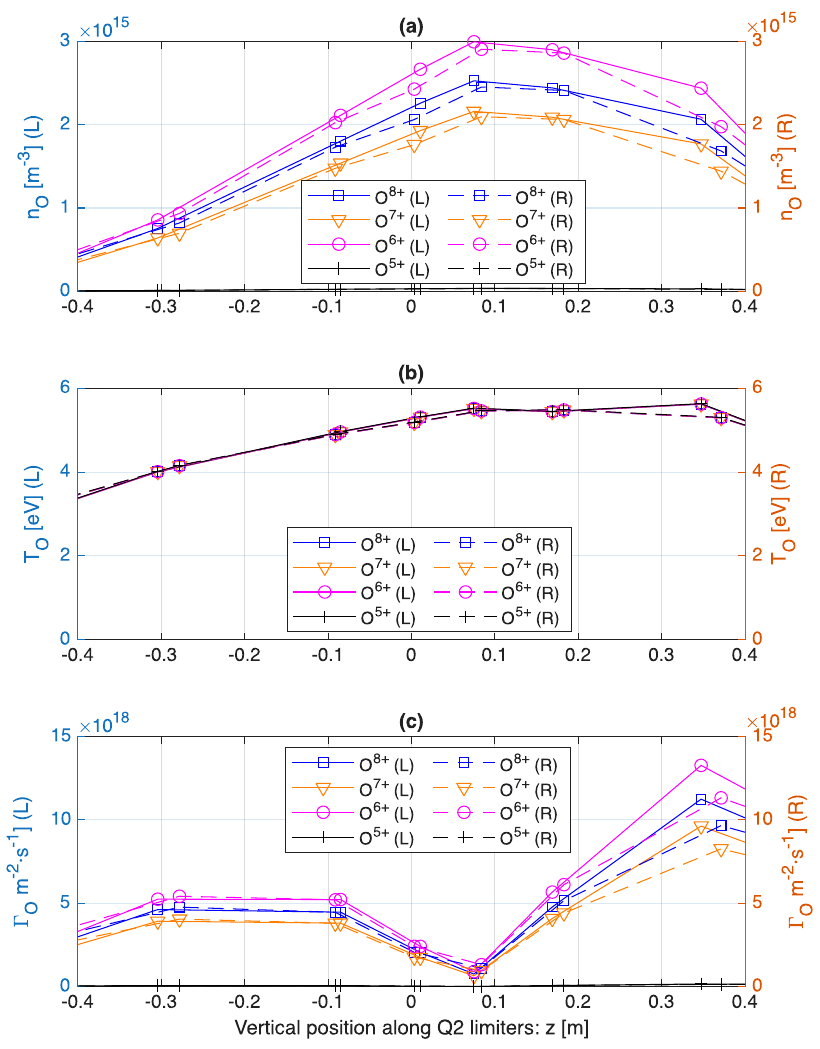}
    \caption{LOS-weighted SolEdge3x limiter profiles for the ICRH phase (\(t = 8\)~s). (a) \(n_{\mathrm{O}^{q+}}\), (b) \(T_{\mathrm{O}^{q+}}\), (c) \(\Gamma_{\mathrm{O}^{q+}}\).}
    \label{fig:limiter_icrh}
\end{figure}

\begin{figure}[h]
    \centering
    \includegraphics[width=\linewidth]{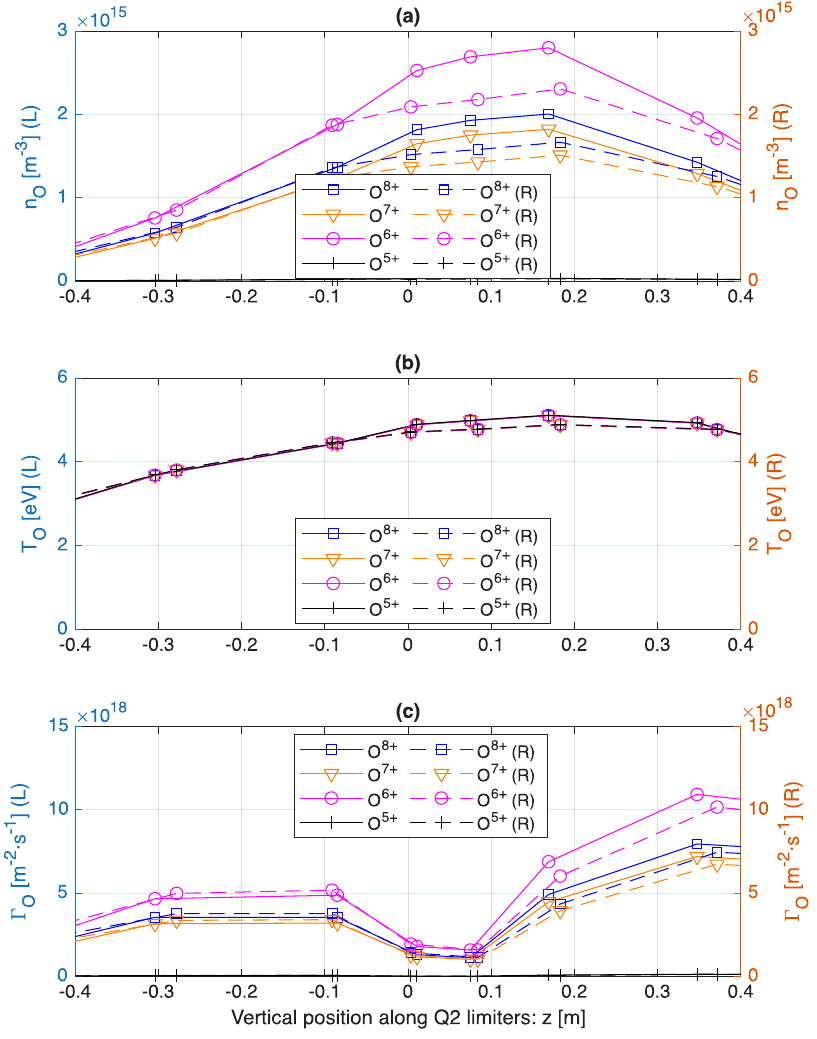}
    \caption{Limiter-resolved SolEdge3x profiles during the ohmic phase (\(t = 3\)~s). Same quantities as Fig.~\ref{fig:limiter_icrh}; RF sheaths are inactive.}
    \label{fig:limiter_ohmic}
\end{figure}

Profiles of \(n_{e^-}\) and \(T_{e^-}\) are mapped along 3D limiter-facing surfaces and then averaged over synthetic line-of-sight (LOS) chords, consistent with both the spectroscopic diagnostic geometry and the STRIPE mesh structure. This approach ensures that simulation inputs remain aligned with the spatial resolution and view factors of experimental diagnostics. Figures~\ref{fig:limiter_icrh} and~\ref{fig:limiter_ohmic} present LOS-weighted vertical profiles  of oxygen ion parameters obtained from SolEdge3x for the four highest charge states (O$^{5+}$ to O$^{8+}$), including density \(n_{\mathrm{O}^{q+}}\), temperature \(T_{\mathrm{O}^{q+}}\), and particle flux \(\Gamma_{\mathrm{O}^{q+}}\) along the left (L) and right (R) Q2 antenna limiters.

\textit{ICRH Phase:}
During ICRH operation, as shown in Figure~\ref{fig:limiter_icrh}, the impurity population near both limiters is dominated by O$^{6+}$, O$^{7+}$, and O$^{8+}$. On the R limiter, \(n_{\mathrm{O}^{6+}}\) peaks near the midplane at approximately \(3.0 \times 10^{15}~\text{m}^{-3}\), while O$^{8+}$ reaches about \(2.5 \times 10^{15}~\text{m}^{-3}\), and O$^{7+}$ slightly less, at approximately \(2.0 \times 10^{15}~\text{m}^{-3}\), with peak values of all three charge states occurring in the upper region (\(z > 0.1~\text{m}\)). The contribution from O$^{5+}$ is negligible throughout the profile.

The vertical distribution of charge states reflects the local \(T_{e^-}\) and the ionization equilibrium. As shown in Figure~\ref{fig:limiter_icrh}, the temperatures \(T_{\mathrm{O}^{q+}}\) vary minimally across charge states and remain within the range of 4--6~eV. The flux profiles \(\Gamma_{\mathrm{O}^{q+}}\) exhibit strong vertical asymmetry: on the R limiter, \(\Gamma_{\mathrm{O}^{6+}}\), \(\Gamma_{\mathrm{O}^{7+}}\), and \(\Gamma_{\mathrm{O}^{8+}}\) all peak sharply in the upper region (\(z \sim 0.3\text{--}0.35~\text{m}\)), with \(\Gamma_{\mathrm{O}^{6+}}\) exceeding \(1.2 \times 10^{19}~\text{m}^{-2}~\text{s}^{-1}\). A clear dip is observed near the midplane (\(z = 0\)), which arises due to tangential magnetic field orientation and stagnation of parallel plasma flow, leading to reduced ion flux at the surface. On the L limiter, the vertical shapes of the flux profiles are similar, but the magnitudes are consistently higher across all charge states. This asymmetry likely originates from the intrinsic geometric asymmetry of the ICRH antenna in the WEST tokamak. However, the toroidal profiles remain largely symmetric due to the axisymmetric wall geometry assumed in {SolEdge3x}. The observed vertical localization of impurity fluxes also coincides with regions of elevated rectified sheath potentials computed using COMSOL, as discussed in Section~\ref{subsec:32}, reinforcing the role of RF sheath rectification as a primary driver of erosion asymmetry.

\textit{Ohmic Phase:}
Under ohmic conditions, the overall impurity density remains comparable to that in the ICRH phase, but the relative charge state distribution shifts slightly as can be infered from Figure~\ref{fig:limiter_ohmic}. Due to the marginally lower oxygen temperatures (\(T_{\mathrm{O}^{q+}} \sim 4.5\text{--}5.0~\text{eV}\)), O$^{6+}$ becomes the most abundant species across both limiters. On the L limiter, peak \(n_{\mathrm{O}^{6+}}\) reaches approximately \(2.8 \times 10^{15}~\text{m}^{-3}\), while O$^{8+}$ and O$^{7+}$ reach about \(2.0 \times 10^{15}~\text{m}^{-3}\) and \(1.8 \times 10^{15}~\text{m}^{-3}\), respectively. These charge states follow similar spatial distributions on the R limiter. O$^{5+}$ remains weakly populated, consistent with limited ionization in the cooler peripheral regions.

Temperatures \(T_{\mathrm{O}^{q+}}\) are relatively flat and symmetric along the vertical axis as shown in Figure~\ref{fig:limiter_ohmic}. The corresponding fluxes \(\Gamma_{\mathrm{O}^{q+}}\) show broad vertical distributions without sharp peaking. Notably, \(\Gamma_{\mathrm{O}^{6+}}\) dominates, with a peak around \(1.2 \times 10^{19}~\text{m}^{-2}~\text{s}^{-1}\), significantly exceeding O$^{7+}$ and O$^{8+}$, which remain below \(0.7 \times 10^{19}~\text{m}^{-2}~\text{s}^{-1}\). The profiles are nearly symmetric between the L and R limiters, reflecting the symmetric boundary conditions in SolEdge3x. This results in uniform impurity exposure and a lower overall erosion intensity compared to the RF-enhanced ICRH scenario.

\subsection{RF-Rectified DC Sheath Potentials from Full-Wave COMSOL Simulations}
\label{subsec:32}

Rectified sheath potentials induced by RF fields are essential boundary inputs to the STRIPE framework and are computed using full-wave electromagnetic simulations in \textit{COMSOL Multiphysics}. These simulations employ the sheath-equivalent dielectric layer method~\cite{Tierens_2024}, which introduces thin layers of spatially varying complex permittivity and conductivity adjacent to PFCs. These material properties are derived from local plasma conditions, enabling the solution of Maxwell’s equations under the cold plasma approximation to self-consistently capture RF field penetration and sheath rectification across realistic 3D geometries.

Input profiles for \(\rm n_{e^-}\) and  \(\rm T_{e^-}\) are taken directly from SolEdge3x simulations, utilizing the wide-grid configuration described in Section~\ref{subsec:31}. Unlike the previous approach in reference \cite{kumar:2025, Tierens_2024} requiring extrapolation to cover the full  RF-antenna components, this implementation benefits from full spatial resolution of the antenna-facing domain—including recessed Faraday Screen bars and antenna sidewalls—facilitating seamless integration into the COMSOL domain.

\begin{figure}[h]
\centering
\includegraphics[width=\linewidth]{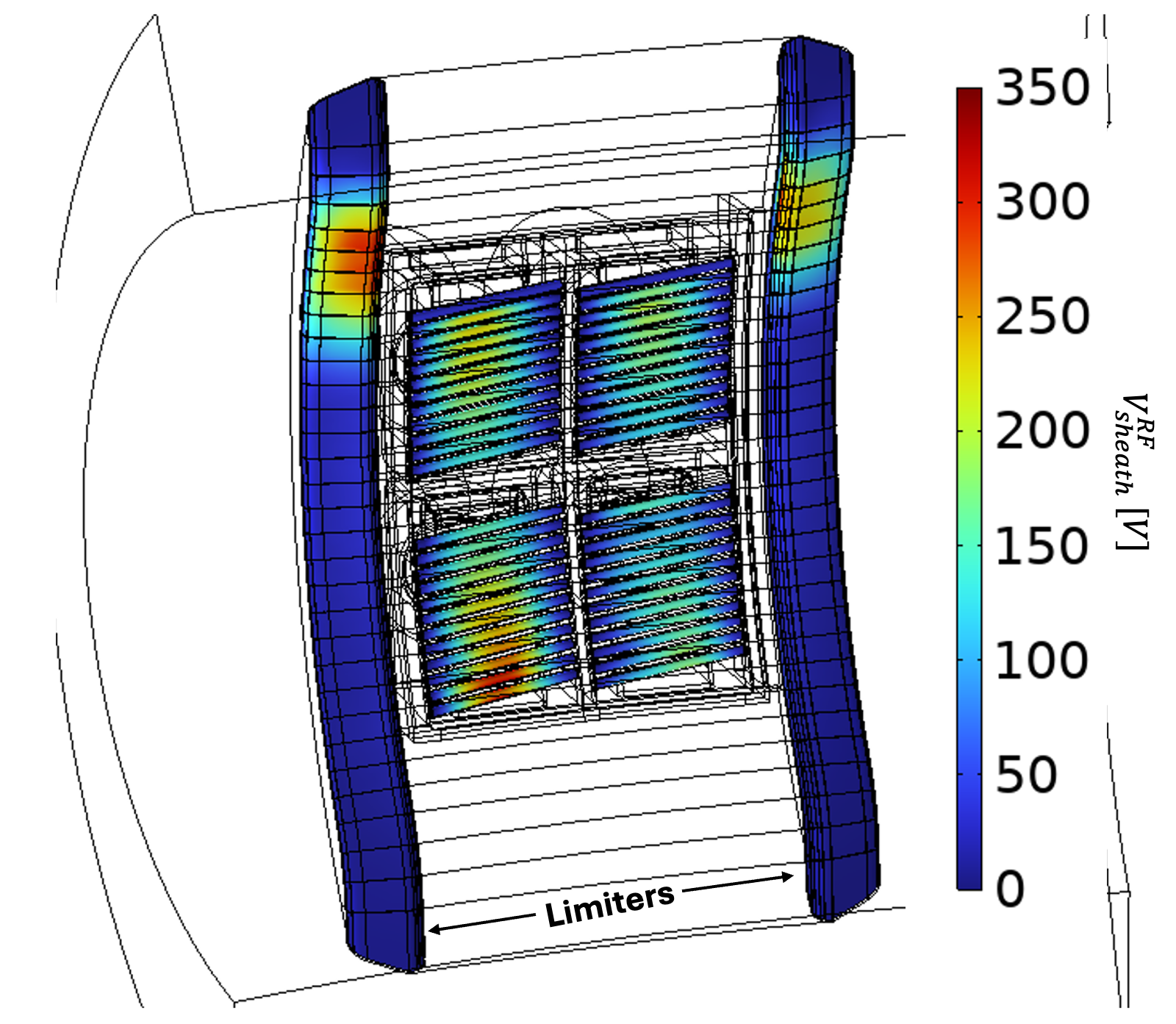}
\caption{3D map of rectified DC sheath potential \(V_{\rm sheath}^{\rm RF}\) computed using COMSOL for WEST discharge \#57877. Peak voltages exceed 250~V and are concentrated near the upper sections antenna limiters' front and sidewalls.}
\label{fig:6}
\end{figure}

Figure~\ref{fig:6} presents the computed 3D distribution of the rectified DC sheath potential \(V_{\rm sheath}^{\rm RF}\) for WEST discharge \#57877. Strong vertical asymmetries are evident, with voltages exceeding 350~V concentrated near upper part of antenna limiters' front and sidewalls. In contrast, lower regions exhibit significantly reduced voltages, typically below 100~V. These variations stem from the interplay between RF strap phasing, magnetic field topology, and nonuniform plasma conditions near the antenna aperture. Importantly, such large DC sheath potentials have also been \textit{directly measured} in WEST using a reciprocating emissive probe, as reported in reference \cite{Diab:2025} which will be used in future for validating STRIPE’s RF sheath predictions.

\begin{figure}[h]
\centering
\includegraphics[width=\linewidth]{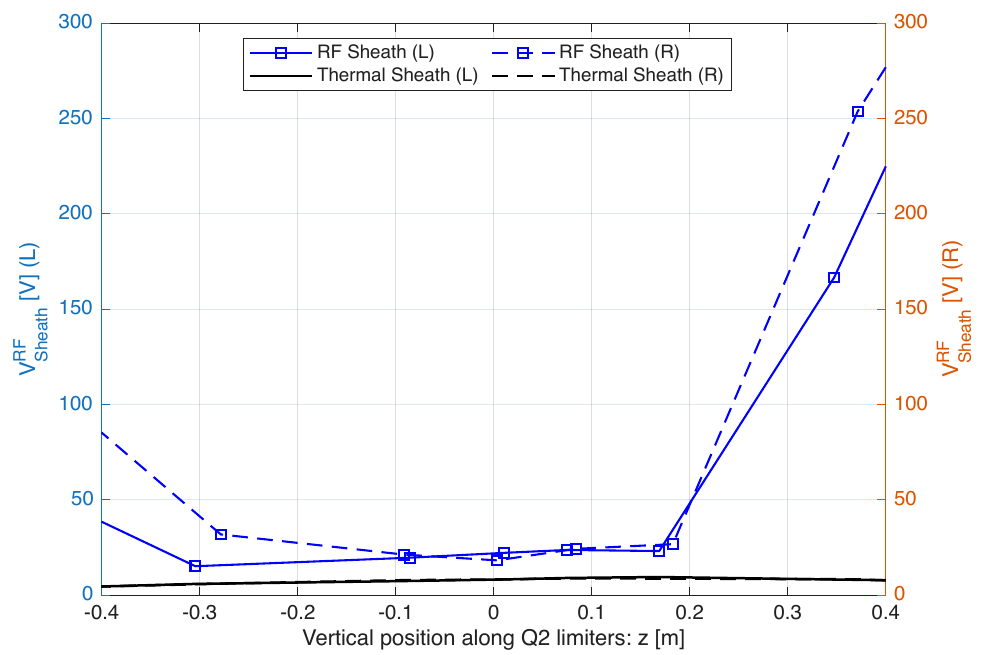}
\caption{Peak \(V_{\rm sheath}^{\rm RF}\) profiles along left and right Q2 limiters during ICRH in WEST \#57877. Voltages exceed 200~V near the top; thermal sheath potentials (black) shown for reference.}
\label{fig:limiter_comsol}
\end{figure}

To further illustrate these trends, Figure~\ref{fig:limiter_comsol} presents 1D vertical profiles of the peak \(V_{\rm sheath}^{\rm RF}\) along the front face of the left and right Q2 antenna limiters during ICRH. Both profiles show pronounced vertical asymmetry, with peak rectified potentials localized near the upper ends, well above the midplane. In contrast, thermal sheath potentials during the ohmic phase are symmetric and centered around the midplane, with peak values near 25~V.

Although $\rm n_{e^-}$ for the discharge \#57877 remains well above the lower-hybrid resonance (LHR) cutoff, future cases closer to the LHR threshold may exhibit stronger slow-wave coupling and intensified sheath rectification—conditions to be investigated in upcoming COMSOL studies. The present model uses a local sheath formulation and does not include DC current equilibration along magnetic field lines. While such currents are treated in SSWICH, recent studies suggest its results align closely with those from Petra-M, which, like STRIPE, employs purely local sheath physics\cite{Urbanczyk:2025}. Notably, due to the magnetic field orientation, field-aligned equilibration would not eliminate toroidal asymmetries.
\subsection{Ion-Surface Interaction and Erosion Modeling}
\label{subsec:33}

STRIPE computes gross W-erosion by combining sheath-modified ion impact dynamics with angle- and energy-resolved sputtering yields. This process involves four key components: electric field modeling, ion trajectory integration, IEAD construction, and effective sputtering yield calculation.

\textit{Electric Field Profile:}
The local sheath electric field \(E_n(r)\) is defined using an analytical  model adapted from Brooks' formulation\cite{Brooks:1990}:

\begin{align}
    E_n(r) = V_{\rm sheath} \bigg[
    & \frac{f_{\rm D}}{2\lambda_{\rm D}} \exp\left(-\frac{r}{2\lambda_{\rm D}}\right) \nonumber \\
    & + \frac{1 - f_{\rm D}}{\rho_i} \exp\left(-\frac{r}{\rho_i}\right)
    \bigg]
    \label{Eq:RFsheath}
\end{align}

where $\rm V_{sheath}$ voltage arising on the antenna structures due the sheath, \(r\) is the distance from the surface, \(\lambda_{\rm D}\) is the Debye length, \(\rho_i\) is the ion gyroradius, and \(f_{\rm D}\) is the fraction of the potential drop across the Debye sheath.

In ICRH phase, $\rm V_{sheath}=V_{sheath}^{RF}$ is taken from COMSOL simulations detailed in subsection \ref{subsec:32}. For the ohmic phase, classical thermal sheath potentials $\rm V_{sheath}=V_{sheath}^{Thermal}$ \cite{stangeby:2000} are calculated as:

\begin{align}
    V_{\rm sheath}^{\rm thermal} &= k_{\rm sheath} T_e \\
    k_{\rm sheath} &= \left| \frac{1}{2} \log\left(2\pi \frac{m_e}{m_i} \right) \right|
    \left(1 + \frac{T_i}{T_e} \right)
    \label{Eq:thermalSheath}
\end{align}

where \(T_e\) and \(T_i\) are the electron and ion temperatures, and \(m_e\), \(m_i\) are their respective masses.

\textit{Ion energy-angle distributions (IEADs):}
IEADs, \(f_i(\mathcal{E}, \theta)\), are computed in STRIPE by tracing impurity ion trajectories from the sheath entrance to the surface using GITR code. At each mesh element, 20,000 test particles, initialized with a half-Maxwellian distribution, launched from the sheath entrance. The local plasma parameters --$\rm n_{e^-}, T_{e^-}, n_{D^+}, T_{D^+}, n_{O^{q+}}, T_{O^{q+}}$, electric ($\rm \mathbf{E}$) and magnetic ($\rm \mathbf{B}$) fields and $\rm V_{sheath}$ -- at each mesh element set the boundary conditions for these simulations.  Their motion is governed by local $\rm \mathbf{E}$ and $\rm \mathbf{B}$ fields, incorporating full gyromotion, \(\mathbf{E} \times \mathbf{B}\) drift, and Coulomb collisions.

Sputtering yields for O sputtering W, \(\rm Y(\mathcal{E}, \theta)(O \rightarrow W)\) or simply \(\rm Y(\mathcal{E}, \theta)\), are precomputed using \textsc{RustBCA} for a grid of monoenergetic, angle-resolved impacts of neutral atomic oxygen on W, spanning ion energies \(\mathcal{E}\) from 10~eV to 5~keV and incidence angles \(\theta\) from 0° to 90°. Like other BCA codes, \textsc{RustBCA} assumes neutral projectiles and does not account for long-range Coulomb interactions or charge-state-dependent stopping. This is a reasonable approximation near material surfaces, where ions are expected to neutralize before impact.

\begin{figure}[htbp]
  \centering
  \includegraphics[width=\linewidth]{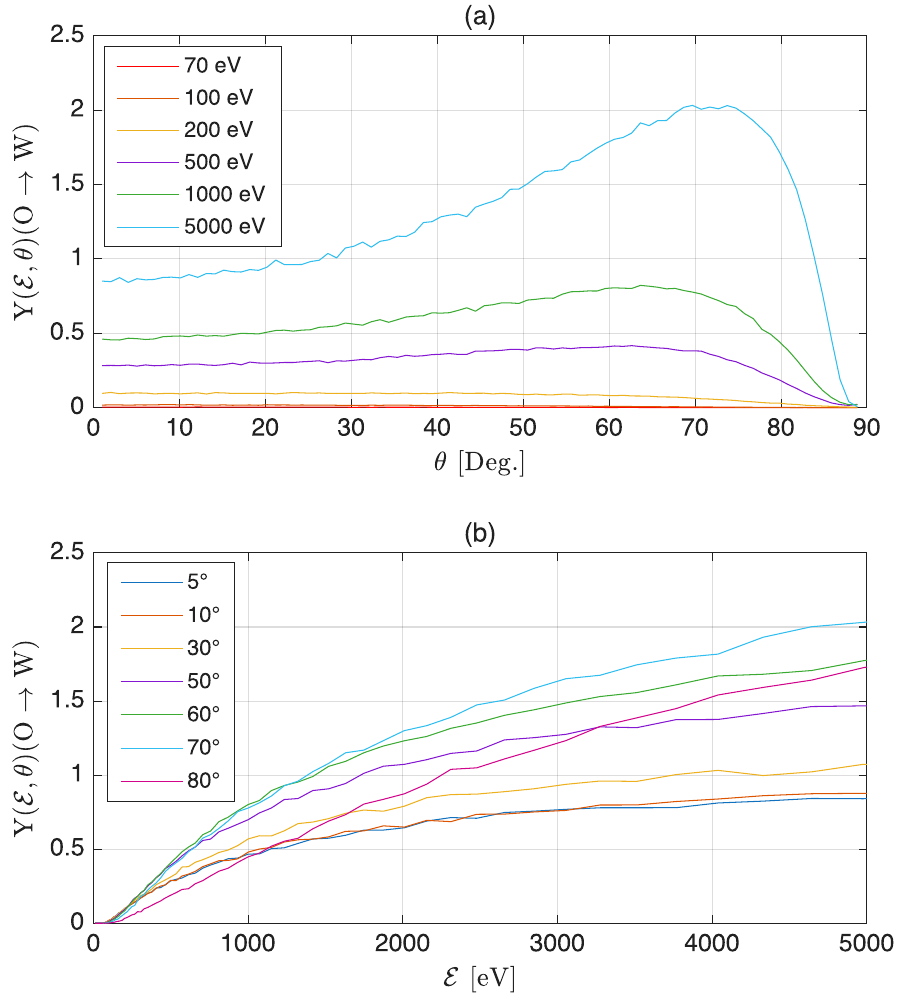}
  \caption{Sputtering yield \(Y(\mathcal{E}, \theta)\) for neutral oxygen atoms incident on W, calculated using \textsc{RustBCA}. 
  (a) Yield vs. incidence angle \(\theta\) for selected energies, \(\mathcal{E}\): 70 eV (red), 100 eV (orange), 200 eV (gold), 500 eV (purple), 1000 eV (green), 5000 eV (blue). 
  (b) Yield vs. energy \(\mathcal{E}\) for selected angles: 5°, 10°, 30°, 50°, 60°, 70°, 80°.}
  \label{fig:rustbca}
\end{figure}

As shown in Figure~\ref{fig:rustbca}, the top panel displays \(\rm Y(\mathcal{E}, \theta)\)  as a function of \(\theta\) for selected  \(\mathcal{E}\). \(\rm Y(\mathcal{E}, \theta)\) increases with \(\theta\), peaking between 70° and 80°, and drops sharply at grazing incidence due to reflection. The bottom panel shows \(\mathcal{E}\) dependence for fixed \(\theta\), where \(\rm Y(\mathcal{E}, \theta)\) increases monotonically with \(\mathcal{E}\) and exceeds 2 atoms/ion at 5~keV for higher \(\theta\). The color-coded legends in Figure~\ref{fig:rustbca} indicate the specific \(\mathcal{E}\) (top) and \(\theta\) (bottom) used here from  the \textsc{RustBCA} simulations, enabling direct interpretation of angular and energetic scaling behavior.

The IEADs from GITR are convolved with these precomputed yields to calculate the local effective sputtering yield:

\begin{align}
Y_{\rm eff}(i) = \int_0^{90^\circ} \!\! \int_{\mathcal{E}_{\rm min}}^{\mathcal{E}_{\rm max}} 
    Y_i(\mathcal{E}, \theta) f_i(\mathcal{E}, \theta)\, d\mathcal{E}\, d\theta
    \label{Eq:spYield}
\end{align}

\textit{Gross Erosion Flux:} Effective yields \(Y_{\rm eff}(i)\) are combined with ion fluxes \(\Gamma_{\rm ions}(i)\) from SolEdge3x to compute the gross W erosion flux $\rm \Gamma_{\rm gross, W}$:

\begin{align}
\Gamma_{\rm gross, W} = \sum_{i=1}^{N} Y_{\rm eff}(i) \cdot \Gamma_{\rm ions}(i)
\label{Eq:grossErosion}
\end{align}

This erosion flux serves as the impurity source for downstream modeling and synthetic diagnostics within STRIPE (see Section~\ref{sec:4}).

\section{Results: W Erosion at the WEST ICRH Antenna Limiters}
\label{sec:4}

This section presents LOS-averaged vertical profiles of \(Y_{\rm eff}\) and  \(\Gamma_{\rm gross, W}\) on the L and R Q2 antenna limiters for WEST discharge \#57877. STRIPE simulations are analyzed for three regimes: (i) ICRH phase with RF sheath, (ii) ICRH phase with thermal sheaths only, and (iii) ohmic phase. Particular attention is given to poloidal and toroidal asymmetries, and to the consistent presence of a midplane dip in erosion and brightness patterns.

\subsection{ICRH Phase: RF Sheath Case}
\begin{figure}[htbp]
    \centering
    \includegraphics[width=\linewidth]{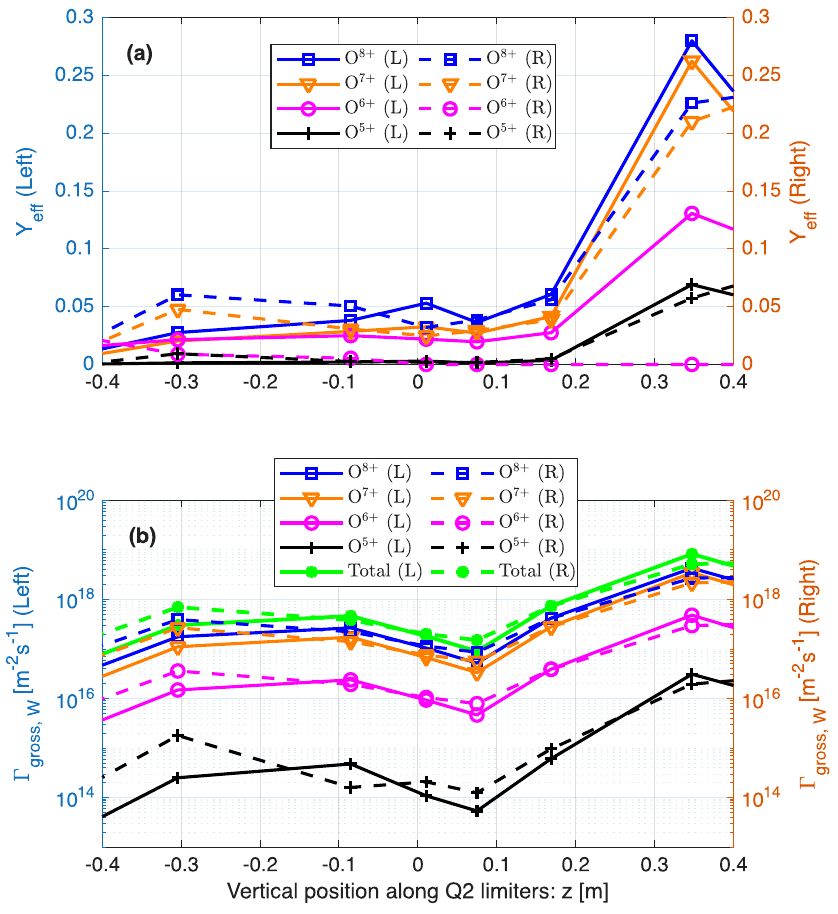}
    \caption{LOS-averaged vertical profiles during ICRH phase with RF sheath: (a) effective sputtering yield \(Y_{\rm eff}\) for O$^{q+}$ (\(q = 5\text{--}8\)); (b) gross erosion flux \(\Gamma_{\rm gross, W}\). Solid lines: L limiter; dashed lines: R limiter. A pronounced midplane dip is observed in both limiters.}
    \label{fig:rf_sheath}
\end{figure}

Figure~\ref{fig:rf_sheath} shows LOS-averaged profiles of \(Y_{\rm eff}\) and \(\Gamma_{\rm gross, W}\) under ICRH conditions with RF sheath potentials included. The RF sheath voltage, computed using full-wave COMSOL simulations (see Section~\ref{subsec:32}), reach peak values exceeding 300~V near the upper limiter surfaces, corresponding to regions with strong field-normal incidence and elevated O$^{q+}$ fluxes.

\textit{Poloidal Trends:} \(Y_{\rm eff}\) for O$^{6+}$, O$^{7+}$  and O$^{8+}$, reaches maximum values near \(z \sim 0.34~\text{m}\) (Figure~\ref{fig:rf_sheath}a), coinciding with the region of highest rectified sheath potential. \(\rm \Gamma_{\rm gross, W}\) exhibits a strongly localized erosion peak at the top of the L limiter, with \(\rm \Gamma_{\rm gross, W}\) exceeding \(5\times 10^{18}~\text{m}^{-2}~\text{s}^{-1}\) as shown in Fig~\ref{fig:rf_sheath}b. This sharp localization indicates a synergy between localized sheath potentials and poloidally asymmetric impurity fluxes. A pronounced dip at the midplane (\(\rm z = 0\)) is also observed, where erosion fluxes drop by more than an order of magnitude. This midplane suppression results from both reduced sheath potential and  more tangential magnetic field orientation, which limit the energy and normal incidence of incoming ions. The combination of these effects leads to a highly non-uniform poloidal erosion profile, with strong localization of W sputtering near the upper limiter surface.

\textit{Toroidal Trends:} The L limiter exhibits higher \(Y_{\rm eff}\) than the R limiter in the upper region (\(z > 0.18~\text{m}\)), where yields from O$^{8+}$ and O$^{7+}$ are strongly localized. In contrast, the R limiter shows slightly elevated yields in the lower region (\(z < 0~\text{m}\)). Similar trends are observed in \(\Gamma_{\rm gross, W}\): the L limiter displays a pronounced peak near the top of the profile, while the R limiter presents a broader distribution with relatively enhanced erosion flux toward the bottom. Although the overall vertical shapes of the profiles are comparable between the two limiters, differences in magnitude and spatial localization result in a clear net toroidal asymmetry. This asymmetry reflects subtle variations in impurity flux distribution (Figure \ref{fig:limiter_icrh}) and pronounced differences in local RF sheath potential exposure (Fig. \ref{fig:limiter_comsol}), consistent with the erosion patterns predicted by STRIPE.

\subsection{ICRH Phase: Thermal Sheath Case}
To isolate the impact of RF rectification, STRIPE simulations were also performed for the ICRH phase using only thermal sheath potentials, with a peak sheath voltage of approximately 25~V.

\begin{figure}[htbp]
    \centering
    \includegraphics[width=\linewidth]{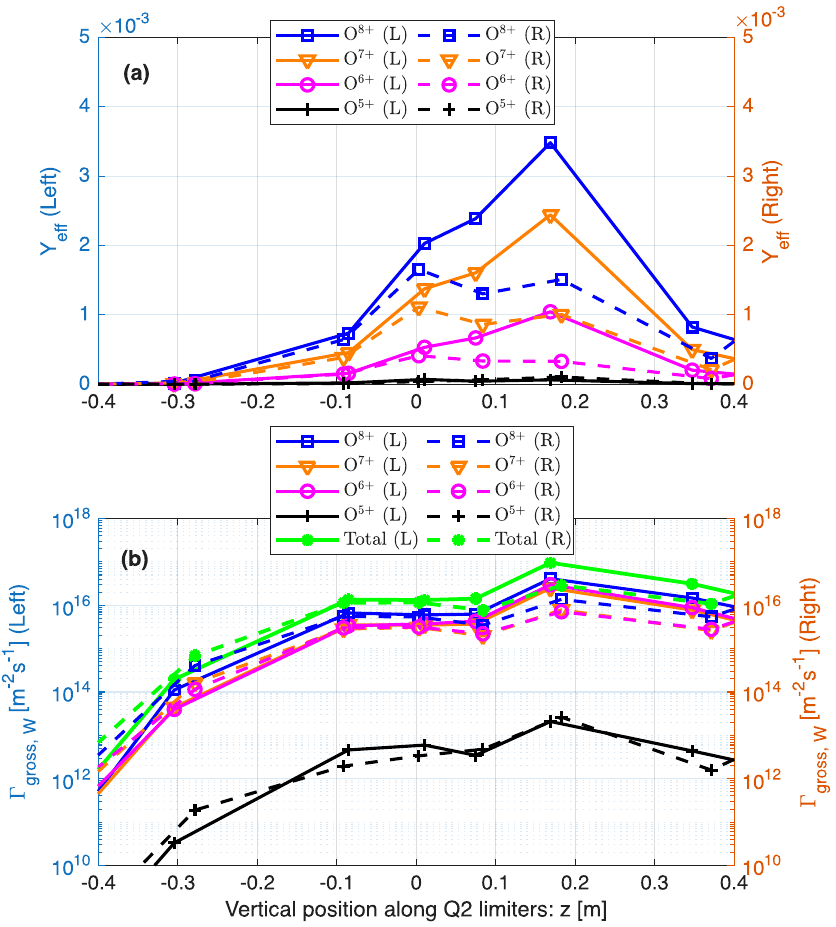}
    \caption{LOS-averaged vertical profiles during ICRH phase with thermal sheath only: (a) \(Y_{\rm eff}\); (b) \(\Gamma_{\rm gross, W}\). The R limiter exhibits a central peak in yield, while the L limiter shows a slight vertical offset.}
    \label{fig:thermal_icrh}
\end{figure}

\textit{Poloidal Trends:} As shown in Figure~\ref{fig:thermal_icrh}(a), \(Y_{\rm eff}\) on the R limiter exhibits a clear central peak at \(\rm z = 0\), consistent with the centrally peaked $\rm V_{sheath}^{Thermal}$ and $\rm n_{O^{q+}}$ in this regime. On the L limiter, the peak in \(Y_{\rm eff}\) is slightly shifted upward to \(\rm z \sim 0.18~\text{m}\), introducing a mild poloidal asymmetry. Across both limiters, the highest yields are observed for O$^{8+}$ and O$^{7+}$, reaching up to \(\rm 3.5 \times 10^{-3}\).

The corresponding \(\Gamma_{\rm gross, W}\), shown in Figure~\ref{fig:thermal_icrh}(b), follows a similarly broad poloidal distribution. On both R \& L limiters, \(\Gamma_{\rm gross, W}\)  increases monotonically toward the upper limiter edge. A \textit{distinct midplane dip}, characteristic of the RF sheath case, is not evident here. The absence of strong vertical gradients reflects the more symmetric nature of the thermal sheath under non-rectified conditions.

\textit{Toroidal Trends:} The L limiter continues to exhibit slightly higher erosion than the R limiter, although the asymmetry is moderate. Overall, the poloidal and toroidal erosion structure is more uniform in this case, reinforcing that the sharp localization seen under RF sheath conditions is driven primarily by sheath rectification effects.

\subsection{Ohmic Phase}

\begin{figure}[htbp]
    \centering
    \includegraphics[width=\linewidth]{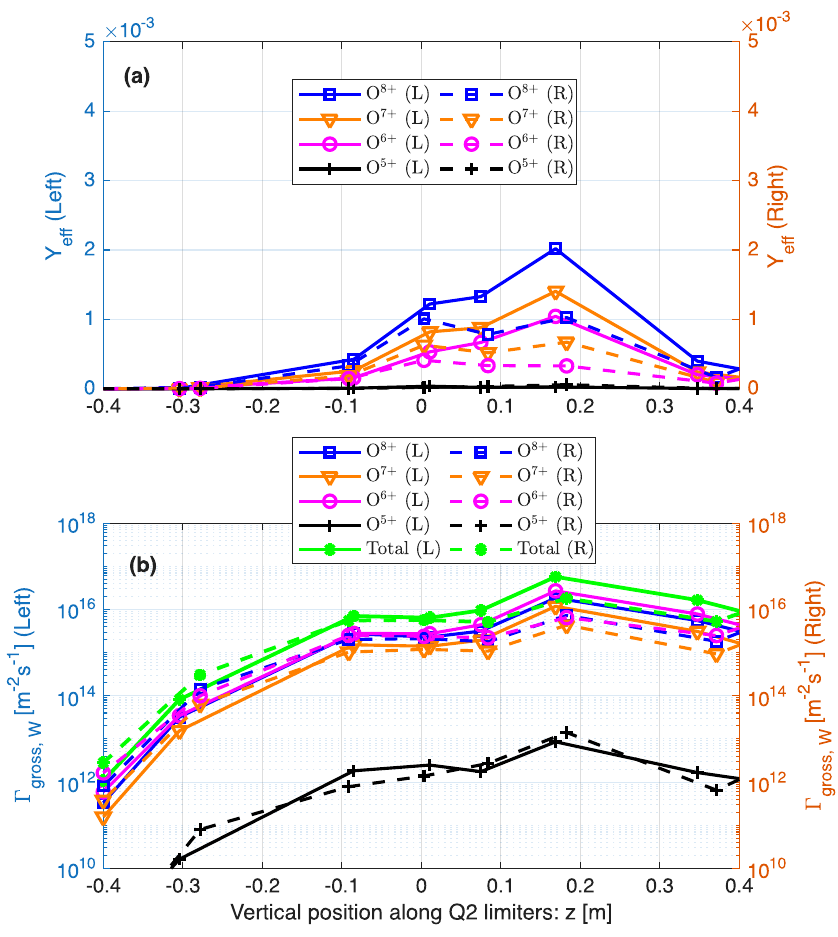}
    \caption{LOS-averaged vertical profiles during the ohmic phase (thermal sheath only): (a) \(Y_{\rm eff}\); (b) \(\Gamma_{\rm gross, W}\). Both limiters exhibit low, symmetric erosion with no midplane suppression.}
    \label{fig:ohmic}
\end{figure}

Erosion predictions during the ohmic phase are shown in Figure~\ref{fig:ohmic}, where RF fields are absent and impurity fluxes are significantly reduced.

\textit{Poloidal Trends:}
The \(Y_{\rm eff}\) and \(\Gamma_{\rm gross, W}\) profiles exhibit distinct poloidal characteristics on the L and R limiters. On the L limiter, \(Y_{\rm eff}\) increases monotonically with vertical position and peaks slightly above the midplane, around \(\rm z \sim 0.18~\text{m}\), indicating a spatial shift from the geometric center. The corresponding \(\rm \Gamma_{\rm gross, W}\) profile rises steadily across the vertical extent and reaches its maximum near the upper end of the limiter. In contrast, the R limiter exhibits a non-monotonic \(\rm Y_{\rm eff}\) profile, with two local maxima at \(\rm z \sim 0~\text{m}\) and \(\rm z \sim 0.18~\text{m}\), and a shallow local minimum near \(\rm z \sim 0.1~\text{m}\). The erosion flux \(\rm \Gamma_{\rm gross, W}\) on the R limiter reflects a similar vertical pattern, with relatively enhanced values toward  the lower  edges. These subtle but measurable poloidal variations influence the local erosion response even under symmetric thermal sheath conditions and reflect differences in local impurity transport and limiter geometry.

\textit{Toroidal Trends:} The L limiter continues to exhibit slightly higher erosion than the R limiter in ohmic phase too. This asymmetry is more in the upper region of the limiter as can also be inferred from the $\rm Y_{eff}$ plot. 

\subsection{Validation Against Experimental Brightness}
\label{subsec:64}

To enable direct comparison with experimental W-I spectroscopy, STRIPE computes synthetic brightness using charge-resolved $\rm \Gamma_{\rm gross, W}$ and \(\rm S/XB\)  coefficients obtained from {ColRadPy}~\cite{curt2019}. The conversion from $\rm \Gamma_{\rm gross, W}$ to brightness $ \rm I_{\phi, W-I}$ is given by:

\begin{equation}
   \rm I_{\phi, W-I} = \frac{1}{A} \int \frac{\Gamma_{\rm gross, W}}{4\pi \, \rm S/XB} \, dA
   \label{eq:4}
\end{equation}

where  \(A\) is the optical collection area, which is set based on a diagnostic radius of 90~mm, following~\cite{Meyer_2018}. This formulation provides a consistent synthetic diagnostic for comparing modeled erosion-driven emission with measured brightness profiles.

\begin{figure}[htbp]
    \centering
    \includegraphics[width=\linewidth]{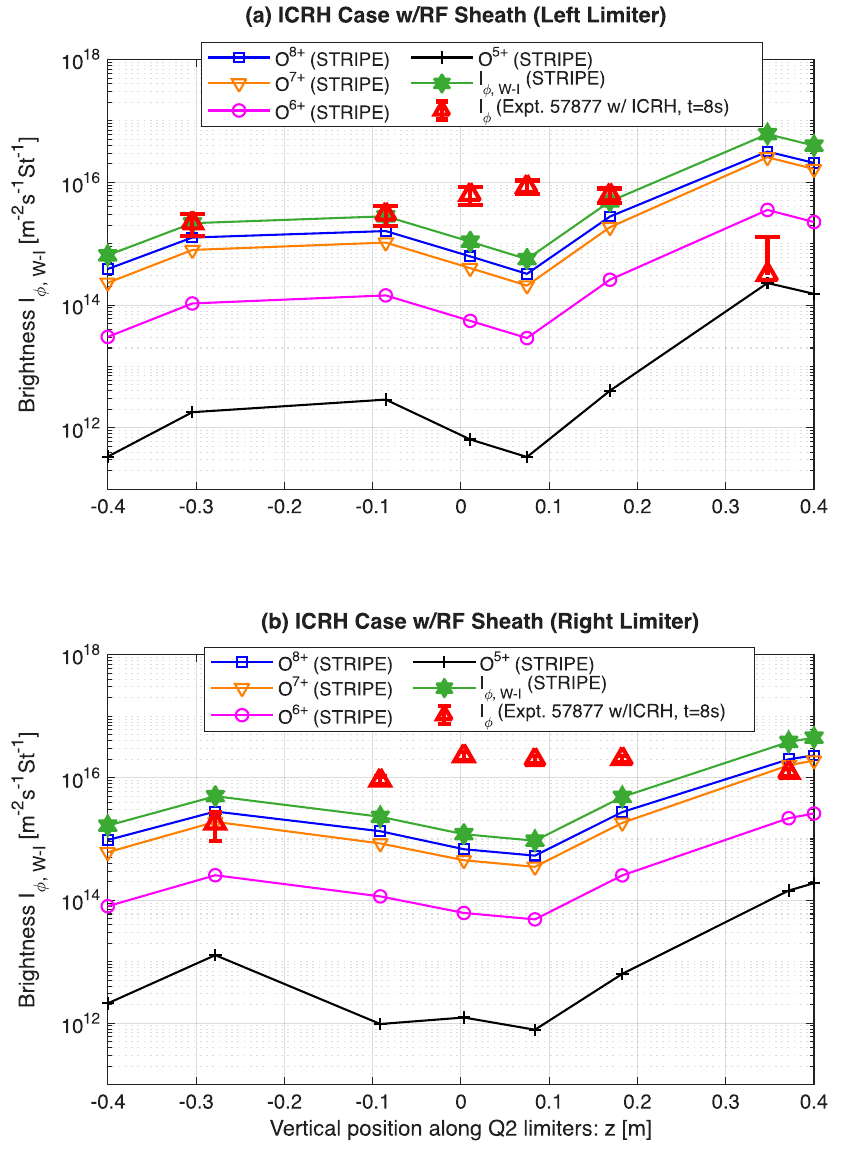}
    \caption{Comparison of STRIPE-simulated and experimental brightness profiles \(\rm I_{\phi, W-I}\) during ICRH with RF sheath. STRIPE predicts upper-limiter localization and L-dominant asymmetry, while the experiment shows midplane peaking and stronger brightness on the R side.}
    \label{fig:brightness_icrh}
\end{figure}

\begin{figure}[htbp]
    \centering
    \includegraphics[width=\linewidth]{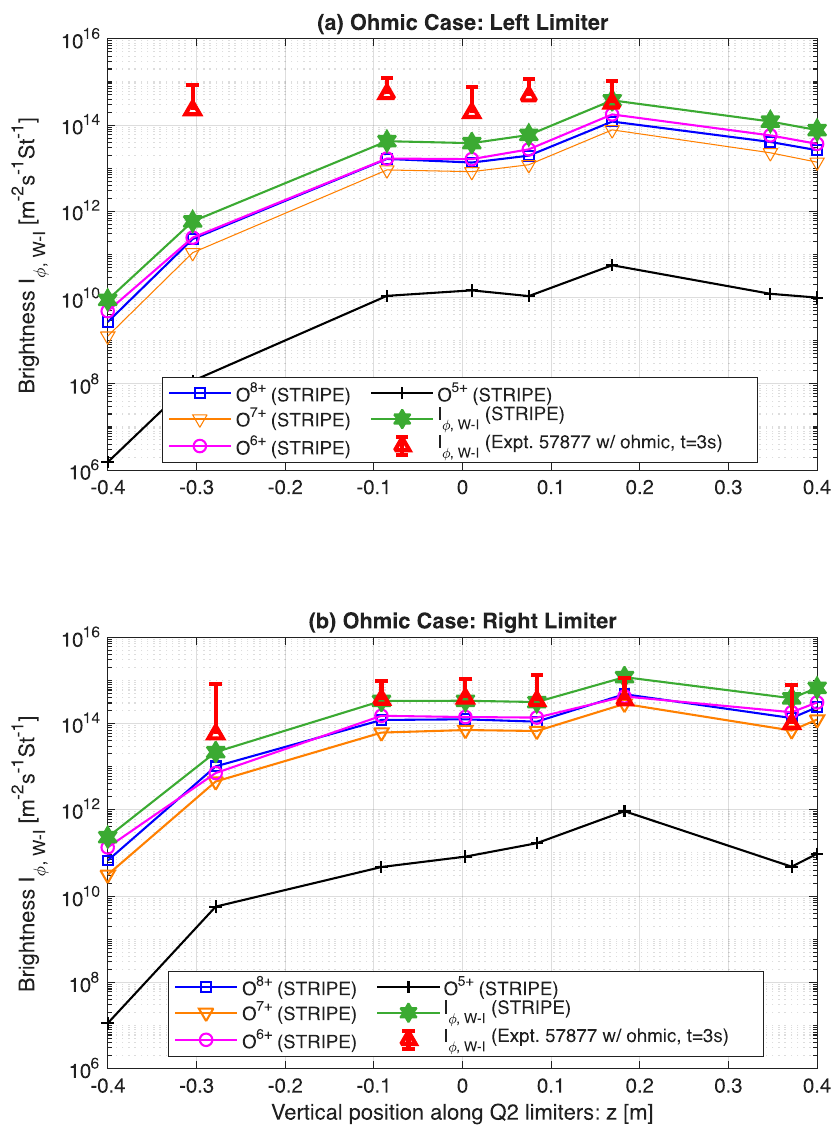}
    \caption{Comparison of STRIPE-simulated and experimental brightness profiles \(\rm I_{\phi, W-I}\) during the ohmic phase. Both model and experiment show vertically uniform emission and toroidal symmetry.}
    \label{fig:brightness_ohmic}
\end{figure}

Figures~\ref{fig:brightness_icrh} and~\ref{fig:brightness_ohmic} compare STRIPE-simulated and experimentally measured W-I 400.9~nm brightness profiles along the Q2 antenna limiters during ICRH and ohmic phases respectively. The comparison addresses spatial trends, asymmetries, and total integrated brightness.

During the ICRH phase (Figure~\ref{fig:brightness_icrh}), the experimental brightness profile peaks near the midplane (\(\rm z = 0\)) and displays a broad, vertically extended distribution along both L and R limiters. In contrast, STRIPE predicts a strongly localized emission peak near the upper limiter region (\(\rm z \sim 0.34~\text{m}\)) and a pronounced dip at the midplane. This spatial discrepancy stems from the modeled localization of RF sheath potentials and impurity fluxes, which diverges from the experimentally observed emission structure. While STRIPE reproduces the overall brightness magnitude reasonably well, it does not capture a significant toroidal asymmetry in W erosion between the L and R limiters. In contrast, experimental measurements show consistently higher brightness on the R limiter, particularly near the midplane, indicating notable mismatches in both toroidal and poloidal distribution. Although STRIPE shows reasonable agreement with the measured profile on the L limiter, it significantly underpredicts emission on the R side—by a factor of 5–10 in the central region. Potential sources of this discrepancy, including RF field phasing, impurity source asymmetry, and reflections in the diagnostic view, are discussed further in Section~\ref{sec:6}.

During the ohmic phase (Figure~\ref{fig:brightness_ohmic}), both STRIPE and experimental measurements exhibit vertically symmetric brightness profiles with a slight upward shift and a clear peak near \(z \sim 0.18~\text{m}\). The modeled \(\rm I_{\phi, W-I}\) closely reproduces the experimental trend, accurately capturing the peak location and the gradual decline toward the lower limiter region. Agreement is particularly strong on the R limiter, where discrepancies remain within a factor of two across the profile. Moreover, no significant toroidal asymmetry is observed in either the model or the experiment during this phase. The absence of RF sheath potentials leads to smoother brightness gradients and reduced overall intensity compared to the ICRH phase. This consistency affirms the accuracy of STRIPE’s charge-resolved impurity transport and thermal sheath modeling in non-RF conditions and further confirms that both poloidal and toroidal asymmetries in W erosion—more pronounced experimentally during ICRH—are primarily driven by RF sheath effects.

A summary of integrated W-I brightness $\rm I_{\phi, total}$, computed from STRIPE  is provided in Table~\ref{tab:brightness_table}. The total modeled brightness agrees with experimental values within 5\% in the ohmic phase and within 30\% during ICRH with RF sheath, supporting STRIPE’s predictive fidelity and confirming the dominant role of O$^{8+}$ and O$^{7+}$ in RF-enhanced sputtering. Contributions from D$^+$ are negligible in all cases.

\begin{table*}
\caption{Simulated charge-resolved W photon brightness \(I_{\phi}\) [photons/s] and comparison with experiment.}
\label{tab:brightness_table}
\centering
\begin{tabular}{lccccccc}
\hline \hline
\textbf{Case} & \(\rm O^{8^+}\) & \(\rm O^{7^+}\) & \(\rm O^{6^+}\) & \(\rm O^{5^+}\) & \(\rm D^+\) & \textbf{Total \(I_{\phi, \rm total}\)} & \textbf{Experiment} \\
\hline
RF (ICRH) & \(2.26 \times 10^{16}\) & \(1.68 \times 10^{16}\) & \(4.02 \times 10^{15}\) & \(1.11 \times 10^{14}\) & \(9.67 \times 10^{10}\) & \(4.35 \times 10^{16}\) & \(3.38 \times 10^{16}\) \\
Thermal (ICRH) & \(1.93 \times 10^{15}\) & \(1.10 \times 10^{15}\) & \(2.58 \times 10^{14}\) & \(8.95 \times 10^{13}\) & 0 & \(3.29 \times 10^{15}\) & -- \\
Thermal (Ohmic) & \(8.05 \times 10^{14}\) & \(4.86 \times 10^{14}\) & \(6.45 \times 10^{13}\) & \(2.66 \times 10^{13}\) & 0 & \(1.39 \times 10^{15}\) & \(1.32 \times 10^{15}\) \\
\hline
\end{tabular}
\end{table*}

\section{Discussion and Summary}
\label{sec:6}

This study applies the STRIPE framework to model impurity-driven W erosion at the WEST ICRH antenna during both ICRH and ohmic phases. STRIPE integrates 2D SolEdge3x plasma backgrounds, full-wave RF sheath potentials from COMSOL, charge-resolved ion transport via GITR, and sputtering yields from RustBCA.

Simulations show that rectified RF sheath potentials significantly amplify gross W erosion, increasing \(\Gamma_{\rm gross, W}\) by more than an order of magnitude compared to thermal sheaths and by over 30× relative to the ohmic case. This enhancement is driven by elevated sheath voltages (up to 300~V) that energize high-charge-state O ions (O$^{6+}$–O$^{8+}$), which dominate the sputtering response. D$^+$ plays a negligible role.

Erosion is poloidally localized near the upper limiter region, consistent with the modeled distribution of RF sheath potentials and impurity fluxes. Toroidally, STRIPE predicts relatively symmetric erosion due to the assumed axisymmetric plasma background. However, experimental W-I brightness during ICRH shows stronger emission on the R limiter and a peak near the midplane, indicating a discrepancy in both toroidal and poloidal behavior. In contrast, during the ohmic phase, both model and experiment exhibit vertically symmetric brightness profiles with peaks near \(z \sim 0.18~\text{m}\), and agreement within a factor of two—supporting the validity of STRIPE in non-RF conditions.

The mismatch in toroidal asymmetry during ICRH may arise from unmodeled 3D plasma features, RF sheath model, or LOS diagnostic limitations. A possible additional factor is a slight radial misalignment between limiters due to mechanical tolerances. Moreover, optical modeling~\cite{Johnson:2024} suggests that W-I brightness measurements may be inflated by reflections from metallic PFCs, contributing up to 95\% of the signal, which may explain the apparent brightness excess measured experimentally.

Nonetheless, the upper-limiter erosion enhancement predicted by STRIPE aligns with trends seen in more recent WEST discharges~\cite{Urbanczyk2025private} and AUG ICRH observations~\cite{BOBKOV2009900}, confirming RF sheath rectification as a primary driver of localized erosion.

STRIPE provides a validated and extensible platform for assessing RF-induced PMI in reactor-relevant environments. Future efforts will include net erosion, re-deposition, and whole-device impurity transport analyses. Planned validation activities include benchmarking COMSOL-predicted sheath potentials against emissive probe measurements, simulating D~I line emission to better constrain input \(n_e\) and \(T_e\), and incorporating mixed light impurity species such as boron, nitrogen, oxygen etc. Methodological developments will also target the inclusion of RF-driven \(\mathbf{E} \times \mathbf{B}\) convection near the antenna---another potential driver of poloidal asymmetry---as well as the incorporation of DC plasma conductivity into the rectification process. These capabilities will be supported through the coupling of 3D {SolEdge3x} with {COMSOL}, enabling more self-consistent modeling of sheath physics and edge plasma dynamics.
\section*{Acknowledgement}
 This work was supported by the U.S. Department of Energy (DOE) Office of Science, through the SciDAC program and utilized resources from the Fusion Energy Division and ORNL Research Cloud under Contract No. DE-AC05-00OR22725. Work at PPPL was supported by DOE Contract No. DE-AC02-09CH11466.
% \section*{References}
% \bibliographystyle{ieeetr}  
\bibliography{west}
\end{document}